\documentclass[11pt,twoside]{article}

\usepackage{asp2004}
\usepackage{epsf}
\usepackage{psfig}
\usepackage{lscape}

\begin{document}
\title{Newly arising problems in the research of SU UMa-type dwarf novae
from VSNET collaborations}   
\author{Daisaku Nogami}   
\affil{Hida Observatory, Kyoto University, Kamitakara, Gifu 506-1314, Japan}    

\begin{abstract} 
Our research on variable objects based on the VSNET collaborations has
achieved much progress in understanding the nature of many kinds of
phenomena.  Many problems have appeared, instead.  Among them, we here
review three newly arising problems in the research of SU UMa-type
dwarf novae: 1) how do EI Psc and V485 Cen evolve as a cataclysmic
variable?, 2) is the early superhump a particular phenomenon for
WZ Sge-type dwarf novae?, and 3) what parameters determine variations
of the superhump period?
\end{abstract}


\vspace{-1cm}
\section{Introduction}   

Cataclysmic variable stars (CVs) are binary systems of a white dwarf
primary, and a late-type main-sequence secondary star.  The surface
gas of the secondary star is transferred to the white dwarf as the
Roche-love overflow via the Lagrangian point (L1).  The transferred
gas forms an accretion disk around the white dwarf, although white
dwarfs in some systems have magnetic fields strong enough to
interrupt the disk formation.

Various kinds of photometric variability are observed in CVs, such as,
flickerings, quasi-periodic oscillations, orbital humps, outbursts,
nova explosions, and so on.  Most of these variabilities are related
to the mass transfer and accretion phenomena.  Dwarf novae are a group
of CVs, which (quasi-)periodically cause large-amplitude outburst
(typically 2-5 mag).

We have been doing research on activities in a variety of variable
objects, such as, dwarf novae, X-ray binaries, GRBs, and so on, with
many amateur/professional astronomers all over the world, mainly by
monitoring many objects, exchanging information, and coordinating
observational campaigns on Variable Star NETwork (VSNET) according to
circumstances \citep{VSNET}.  Much progress in understanding the
nature of many kinds of variable objects and phenomena has been
achieved.  Many problems have appeared, instead.  Among them, we will
review three newly arising problems in the research of SU UMa-type
dwarf novae: 1) how do EI Psc and V485 Cen evolve as a cataclysmic
variable?, 2) is the early superhump a particular phenomenon for
WZ Sge-type dwarf novae?, and 3) what parameters determine variations
of the superhump period ($P_{\rm sh}$)?

\section{Three new problems}

\subsection{How do EI Psc and V485 Cen evolve as a cataclysmic variable?}

This problem was highlighted by Uemura et al. (2002a, b).  EI Psc is a
counterpart of an X-ray source 1RXS J232953.9+062814.  The dwarf nova
nature of this object was proved by optical spectra showing hydrogen
and helium emission lines and TiO absorption bands \citep{hu98j2329}.
The first outburst of EI Psc was reported to VSNET by P. Schmeer on
2001 November 3.  An observation campaign was promptly coordinated by
Uemura et al.  They detected superhumps during this outburst, revealing
that EI Psc is an SU UMa star.

The most interesting point is that EI Psc has a very short orbital
and superhump period of 64.2 min and 66 min \citep{uem02j2329letter,
tho02j2329}, respectively.  This orbital period breaks the so-called
``observed period minimum'' of 78 min.  Moreover, its relatively bright
quiescence magnitude and spectrum in quiescence suggest that this star
has a relatively high rate of the mass transfer from a mid K-type
secondary star.  This does not agree with the standard evolution theory
telling that dwarf novae around the period minimum have a very low-mass
secondary of the late M- or early L-type, and the mass transfer rate is
quite low.  With the analogous system of V485 Cen (see Olech 1997, and
references therein), which was somewhat ignored due to its uniqueness,
these objects establish the first subpopulation in hydrogen-rich
cataclysmic variables below the period minimum.

Podsiadlowski et al. (2003) predicted that a cataclysmic variable whose
secondary star has a hydrogen-exhausted core has a shorter period
minimum and the secondary is more luminous than in a CV having a normal
secondary star.  The peculiarity of EI Psc and V485 Cen is generally
consistent with this view.  As suggested by Podsiadlowski et al. (2003),
the group of EI Psc and V485 Cen may be on the evolution path to the
double-degenerated AM CVn-type binaries \citep{uem02j2329}.

SDSS J013701.06$-$091234.9, the most recently discovered SU UMa-type
star (Imada et al. in this volume; Imada et al. 2006), seems to have
intermediate properties between the normal SU UMa stars and the
EI Psc/V485 Cen group in terms of the orbital period, mass transfer
rate, type of the secondary star.  This may be a bridging object between
these two classes.

\subsection{Is the early superhump a particular phenomenon for WZ Sge-type
dwarf novae?}

WZ Sge stars are a small group of enigmatic SU UMa-type dwarf novae (see
Kato et al. 2001, and references therein).
These stars have the following common outburst properties: 1) very long
recurrence cycles of the outburst (years to decades), 2) large outburst
amplitudes over 6 mag, 3) long outburst durations (including the
rebrightening phase and long fading tail) up to 100 days, and 4) no (or
only few) normal outburst between successive two superoutbursts.  While
the outburst behavior of normal dwarf novae are well explained by the
thermal-tidal disk instability model, that of WZ Sge stars still remains
a big challenge (for a review, Osaki 1996).

However, another common phenomenon, called `early superhumps', has been
recently recognized.  The early superhumps are observed before emergence
of the (normal) superhumps, and have doubly peaked shapes while the
(normal) superhumps have singly peaked shapes.  The period of the early
superhump is quite close to the orbital period [Note that the
early-superhump period observed during the 2001 superoutburst in WZ Sge
was shorter by 0.05\% than the orbital period.  This difference was
small, but significantly larger than the error.].  Early superhumps have
been observed all the four definite WZ Sge stars intensively observed
from the early phase of superoutbursts.  They are WZ Sge itself
\citep{ish02wzsgeletter, pat02wzsge}, AL Com \citep{kat96alcom,
ish02wzsgeletter}, HV Vir \citep{kat01hvvir,
ish03hvvir}, and EG Cnc \citep{mat98egcnc, kat04egcnc}.

There is another SU UMa-type dwarf nova where early superhumps were
observed.  It is RZ Leo whose SU UMa nature was revealed during the
2000-2001 superoutburst \citep{ish01rzleo}.  The following features of
RZ Leo are common to the canonical WZ Sge stars: 1) the recent low
frequency of the outburst (the 2000 superoutburst is only one outburst
caught in 1995-2005), 2) the large outburst amplitude of $\sim$6 mag,
and 3) the long outburst duration of $>$40 days.  RZ Leo is, however,
different from other WZ Sge stars in the three points: 1) the `long'
orbital period of 0.0765(2) d \citep{men01rzleo} (cf. $\sim$0.058 d in
other WZ Sge stars), 2) the large excess (3.3\%) of the superhump
period to the orbital period (cf. $\sim$1\% in other WZ Sge stars), and
3) the existence of a period of a `short' outburst cycle (one outburst
every year in 1987-1990).  The superhump excess has a tight relation
with the mass ratio ($q = M_{\rm 2}/M_{\rm 1}$), and the large
superhump excess of RZ Leo suggests the mass ratio of RZ Leo
($\sim$0.14) to be about two times larger than that of other WZ Sge
stars.

The current low outburst frequency and behavior during the 2000-2001
superoutburst support RZ Leo to be a WZ Sge-type dwarf nova.  The
variation of the outburst frequency implies a change of the mass
transfer rate.  RZ Leo may be presently in the `WZ Sge phase' with
a low mass transfer rate.  If this is the case, it implies that the
orbital period and mass ratio are only weakly related to the WZ Sge
phenomena, and that the mass transfer rate is the unique parameter
to distinguishes the normal SU UMa stars and the WZ Sge systems.

Judging from the observations available at this time, there is no
negative evidence for that the early superhump is a particular
phenomenon for WZ Sge-type dwarf novae.  The samples, however,
must not be enough to conclude this problem.  We should continue
to monitor RZ Leo and to start coordinated observations of
WZ~Sge/SU~UMa systems in outburst as early as possible.

\subsection{What parameters determine variations of the superhump
period?}

The superhump period had been considered to gradually decrease, or
at least remain constant, during one superoutburst, till 1996.
However, since the discovery of a $P_{\rm sh}$ increase in the 1996
superoutburst in SW UMa \citep{sem97swuma, nog98swuma}, the same
phenomenon have been observed in some dwarf novae (Imada et al. 2005;
Uemura et al. 2005, and references therein).

Figure 5 in \citep{ima05gocom} summarizes the time derivatives
($P_{\rm dot} \equiv \dot{P}_{\rm sh}/P_{\rm sh}$) of the superhump
period ever measured.  In this figure, we can see a trend that dwarf
novae having a shorter $P_{\rm sh}$ have a high probability that the
superhump period is observed to increase.

Based on Osaki and Meyer (2003), this phenomenon is interpreted in the
following way.  In the SU UMa-type dwarf novae having a relatively
long orbital period and a relatively high mass transfer rate, the 3:1
resonance radius in the accretion disk is very close to the tidal
truncation radius, and the mass is not much accumulated by
a superoutburst due to frequent occurrence of the outburst.  Then,
the eccentric waves arising at the 3:1 resonance radius can not
spread outwards, but do inwards.  The superhump period thus decrease.
On the other hand, in short period systems infrequently causing
outbursts, the 3:1 resonance radius is significantly smaller than
the tidal truncation radius, and the mass is much accumulated in
the accretion disk.  Then, enough mass spreads over the 3:1
resonance radius at the onset of a superoutburst, and the eccentric
waves can propagate outwards.  It is thus observed that the
superhump period increases.

Uemura et al. (2005) observed in TV Crv that the superhump period
increased during the 2001 superoutburst with no precursor, and the
superhumps period remained almost constant during the 2004
superoutburst with a precursor.  The existence of the precursor
in the 2004 superoutburst suggests that the mass accumulated during
quiescence was smaller before the 2004 superoutburst than before
the 2001 superoutburst.  These observations support the
interpretation described above.

In conclusion, the orbital period and mass transfer rate should be
important parameters for the variation of the superhump-period
derivative.  However, it has been recently observed that the
superhump period remained constant, or might increase even in some
SU UMa stars having relatively long orbital periods.  In addition,
some stars show interchanges of the increase/decrease trend of the
superhump period (e.g. TT Boo, see Olech et al. 2004).  There may
be other parameter, for instance, the mass ratio, related to the
variations of the superhump period.  We still need more
observations to fully solve this problem.


\vspace{5mm}
We are sincerely thankful to all the VSNET collaborators.  This work is
supported by Grants-in-Aid from the Ministry of Education, Culture,
Sports, Science and Technology (MEXT) (Nos. 16340057, and 17740105).


\end{document}